%% file: main.tex
\documentclass[10pt,
prd,
twocolumn,
nofootinbib,
noshowkeys,
noshowpacs,
superscriptaddress,
floatfix
]{revtex4-2}
\usepackage{amsmath,amsfonts,amsthm,amssymb}
\usepackage[normalem]{ulem}
\usepackage[dvips]{graphics,graphicx}
\usepackage[usenames,dvipsnames]{color}
\definecolor{darkblue}{RGB}{0,0,196}
\definecolor{darkgreen}{RGB}{0,120,0}
\usepackage[colorlinks=true,linktocpage=true,linkcolor=darkblue,citecolor=red,urlcolor=darkblue]{hyperref}
\usepackage{cancel}
\usepackage{bbold}
\usepackage{subfigure}\usepackage{stix}
\usepackage{multirow}
\usepackage{longtable}
\usepackage{color}
\usepackage[normalem]{ulem}
\usepackage{hyperref}
\usepackage{bigints}
\usepackage{xparse}
\usepackage{physics}
\usepackage{verbatim}
\usepackage{minibox}
\usepackage{comment}
\usepackage{appendix}
\usepackage{scalerel}
\usepackage{marginnote}
\usepackage{graphicx}
\usepackage[nice]{nicefrac}
\usepackage{soul}
\usepackage{bm}
\usepackage{combelow}

\usepackage{stackengine}

\def\HP{\hphantom{\alpha}} 
\usepackage{amsmath}
\usepackage{hepunits}
\include{commands}

\begin{document}
\preprint{}
 
    \title{
    Spin waves in spin hydrodynamics} 
	\author{Victor E. Ambru\cb{s}}
	\email{victor.ambrus@e-uvt.ro}
    \affiliation{
Institut f\"ur Theoretische Physik, Johann Wolfgang Goethe-Universit\"at, Max-von-Laue-Strasse 1, D-60438 Frankfurt am Main, Germany}%
\affiliation{Department of Physics, West University of Timi\cb{s}oara, \\
Bd.~Vasile P\^arvan 4, Timi\cb{s}oara 300223, Romania}

	\author{Radoslaw Ryblewski}
\email{radoslaw.ryblewski@ifj.edu.pl}
	\author{Rajeev Singh}
\email{rajeev.singh@ifj.edu.pl}
\affiliation{Institute of Nuclear Physics Polish Academy of Sciences, PL 31-342 Krak\'ow, Poland}%
	\date{\today} 
	\bigskip
\begin{abstract}
The propagation properties of 
spin degrees of freedom are analyzed
in the framework of relativistic hydrodynamics with spin based on the de Groot--van Leeuwen--van Weert definitions of the energy-momentum and spin tensors.
We derive the analytical expression for the
spin wave velocity for arbitrary statistics
and show that it goes to half the speed of light in the ultrarelativistic limit.
We find that only the transverse degrees 
of freedom propagate, analogously to electromagnetic waves.
Finally, we consider the effect of dissipative 
corrections and calculate the damping coefficients for the
case of Maxwell-J\"uttner statistics.
\end{abstract}
     
\date{\today}


	
	\maketitle
\section{Introduction}
Recent spin polarization measurements of $\Lambda (\Bar{\Lambda})$ hyperons~\cite{STAR:2017ckg,Adam:2018ivw,Niida:2018hfw,STAR:2019erd,ALICE:2019aid,ALICE:2019onw,STAR:2021beb,ALICE:2021pzu} have 
sparked a huge interest in the heavy-ion physics community.
In this context, many theoretical studies have been performed referring to spin-orbit coupling \cite{Liang:2004ph,Liang:2004xn,Gao:2007bc,Chen:2008wh}.
Fundamentally, the polarization of particles with spin can be induced through the spin-orbit coupling implied by the Dirac equation \cite{Itzykson:1980rh,Peskin:1995ev}. Starting with the works by Vilenkin in the 1980s \cite{Vilenkin:1980zv}, it is now understood that a gas of Dirac particles in rigid motion develops a flow of chirality along the vorticity direction~\cite{Kharzeev:2015znc}.
Due to its close relation with the axial anomaly, the flow of chirality due to either background vorticity or electromagnetic fields is understood as ``anomalous transport'' \cite{Kharzeev:2015znc}.
Attempts to incorporate such effects dynamically have lead to the development of the so-called hydrodynamics with triangle anomalies \cite{Son:2009tf}. 
While the persistent polarization of massless particles can be modeled via an axial chemical potential, such an approach is not justified for massive particles, where the conservation of the axial current is explicitly broken (alternatively the helical chemical potential may be used, as discussed in Refs.~\cite{Ambrus:2019ayb,Ambrus:2019khr,Ambrus:2020oiw}).

Various models based on the thermodynamic equilibrium of spin degrees of freedom~\cite{Becattini:2017gcx,Becattini:2021suc,Becattini:2021iol,Fu:2021pok} have shown good agreement with experimental data of spin polarization, for recent reviews and papers on this topic see, e.g.,
Refs.~\cite{Wu:2019eyi,Florkowski:2018fap,Weickgenannt:2020aaf,Speranza:2020ilk,Liu:2020ymh,Becattini:2020ngo,lisa2021,Hidaka:2022dmn}.
Nevertheless, the differential measurements of polarization~\cite{STAR:2019erd,ALICE:2021pzu} lack a clear explanation.
This led to the idea of including spin degrees of freedom in standard hydrodynamics, first proposed in Refs.~\cite{Florkowski:2017ruc,Florkowski:2017dyn} based on the definitions of the energy-momentum and spin tensors introduced by de Groot, van Leeuwen, and van Weert (GLW)~\cite{DeGroot:1980dk}.
For recent studies on this formalism see Refs.~\cite{Florkowski:2018ahw,Florkowski:2019qdp,Bhadury:2020puc,Singh:2020rht,Singh:2021man,Florkowski:2021wvk}.

In this work we consider the propagation properties of linear perturbations~\cite{cercignani2002relativistic,rezzolla2013relativistic,Monnai:2014qaa,Ambrus:2017keg,Hongo:2021ona} in the framework of the perfect-fluid spin hydrodynamics~\cite{Florkowski:2018fap,Florkowski:2018ahw},
for other similar studies using the effective action approach, see Refs.~\cite{Montenegro:2017rbu,Montenegro:2017lvf,Montenegro:2020paq}.
At the level of the spin conservation 
equation, we find that the spin degrees of freedom are
decoupled from the background fluid, and therefore 
their wave spectrum can be analyzed separately from the fluid degrees of freedom. Conversely, the fluid degrees of freedom 
are also decoupled from the spin ones leading to the well-known sound waves~\cite{cercignani2002relativistic,rezzolla2013relativistic,Monnai:2014qaa,Ambrus:2017keg}. 
In this study, we consider a linearized expression for the spin tensor in which quadratic or higher order terms are neglected.
For this reason, our results are strictly valid only for the case of propagation through an unpolarized background. In this case, we obtain a general analytic expression for the spin wave velocity, which we apply to the case of both Maxwell-J\"uttner (MJ) and Fermi-Dirac (FD) statistics. In addition, we also derive the relativistic and nonrelativistic limits of the spin wave velocity $c_{\rm spin}$. In both cases, $c_{\rm spin} = c/2$ in the ultrarelativistic limit, with $c$ being the speed of light.
The spin degrees of freedom can be split into an electric part, $C_{\boldsymbol{\kappa}}$, and a magnetic part, $C_{\boldsymbol{\omega}}$, in analogy with the electromagnetic degrees of freedom. We find that the degrees of freedom corresponding to the longitudinal direction (which is parallel to the wave vector $\mathbf{k}$) do not propagate, while the four transverse ones support the usual linear or circular polarization, in perfect analogy to the case of electromagnetic waves \cite{Jackson:1998}.
Finally, using the dissipative spin tensor derived in Refs.~\cite{Bhadury:2020puc,Bhadury:2020cop} for the case of the ideal MJ gas, we discuss the effect of dissipative corrections leading to exponential damping of both the transverse and the longitudinal components.
\par The paper is organized as follows. We begin with a brief review of the formalism of spin hydrodynamics in Sec.~\ref{sec:spinhydro}. Then, in Sec.~\ref{sec:wave_analysis}, we study the propagation of perturbations in the spin polarization components and present the spin wave solutions. Subsequently, in Sec.~\ref{sec:diss}, we analyze the effects of dissipation on the spin wave propagation. Finally, we conclude in Sec.~\ref{sec:conclusion}. Technical details about the spin tensor for arbitrary statistics, the ideal gas, and the FD gas can be found in Appendices \ref{app:spin}, \ref{app:ideal}, and 
\ref{app:FD}, respectively.

In this work, we use the convention of the Minkowski metric $g_{\mu\nu} = \hbox{diag}(+1,-1,-1,-1)$, while the dot product of two four-vectors $a^{\alpha}$ and $b^{\alpha}$
reads $a \cdot b =a^{\alpha}b_{\alpha}= g_{\alpha \beta} a^\alpha b^\beta = a^0 b^0 - \av \cdot \bv$, where boldface indicates three-vectors. For the  Levi-Civita tensor $\epsilon^{\alpha\beta\gamma\delta}$ we use the convention $\epsilon^{txyz} = +1$. We denote the antisymmetrization by a pair of square brackets as $M_{[\mu \nu]} = \frac{1}{2}\left(M_{\mu\nu} - M_{\nu\mu} \right)$. Moreover, we assume natural (Planck) units, {\it i.e.} $c = \hbar = k_B~=1$ (unless stated explicitly).
%
\section{Perfect-fluid spin hydrodynamics}
\label{sec:spinhydro}
In this section, we briefly review the hydrodynamic framework based on the GLW definitions of energy-momentum and spin tensors for the case of spin-$\frac{1}{2}$ particles with mass $m$~\cite{Florkowski:2018ahw,Florkowski:2018fap}.
In this framework, the spin effects are assumed to be small so that the conservation laws for charge, energy, and momentum are independent of the spin tensor. The spin effects arise only from the conservation of angular momentum~\cite{Florkowski:2018ahw,Florkowski:2018fap}. 
The conservation laws of baryon current and energy-momentum tensor are defined, respectively,
as~\cite{Florkowski:2017ruc,Florkowski:2018ahw,Florkowski:2018fap}
\ba
\p_\alpha N^\alpha(x)  = 0\,, \quad \p_\b T^{\a\b}(x) = 0\,,
\lab{Ncon}
\ea
where the baryon current, $N^\alpha$, and the energy-momentum tensor, $T^{\alpha\beta}$, are of the form~\cite{Florkowski:2017ruc}
\ba
N^\alpha = {\cal N} U^\alpha\,, \quad 
T^{\a\b} = {\cal E} U^\a U^\b - 
{\cal P} ~\Delta^{\a\b}\,,
\lab{Nmu}
\ea
with ${\cal N}$, ${\cal E}$, and ${\cal P}$ being the baryon charge density, energy density, and pressure respectively. The fluid four-velocity is denoted by $U^{\mu}$ and  $\Delta^{\a\b} = g^{\a\b} - U^\a U^\b$ is the projector onto  the hypersurface orthogonal to $U^{\mu}$.

Due to the symmetric nature of the energy-momentum tensor (\ref{Nmu}), the conservation of total angular momentum dictates the separate conservation of spin~\cite{Florkowski:2018ahw}
\beq
\p_\a S^{\a , \beta \gamma }(x)&=& 0.
\label{eq:SGLWcon}
\eeq
Violations of the above conservation equation 
can be induced through quantum effects such as nonlocal
collisions
\cite{Hidaka:2018ekt,Weickgenannt:2020aaf,Yang:2020hri,Wang:2020pej,Weickgenannt:2021cuo}, leading most likely to a relaxation of the spin
polarization tensor $\omega^{\mu\nu}$
\eqref{spinpol1} towards the local 
thermal vorticity.
Since the exact form of this relaxation equation is not 
known yet, we do not consider such effects in this analysis.
To the leading order in $\omega^{\beta\gamma}$,
the spin tensor can be decomposed as 
\cite{Florkowski:2018ahw,Florkowski:2018fap,Florkowski:2021wvk}
\begin{subequations}\label{eq:S}
\beq
S^{\alpha , \beta \gamma }
&=&  S^{\alpha , \beta \gamma }_{\rm ph} + S^{\a, \b\g}_{\Delta},
\label{eq:SGLW}
\eeq
where the phenomenological $S^{\alpha , \beta \gamma }_{\rm ph}$
and the auxiliary $S^{\a, \b\g}_{\Delta}$ contributions are given
by \cite{Florkowski:2017ruc,Florkowski:2021wvk}
\beq
S^{\alpha , \beta \gamma }_{\rm ph}
&=&  (\mathcal{A}_1 + \mathcal{A}_3) U^\alpha \omega^{\beta\gamma},\label{Spheno}\\
S^{\a, \b\g}_{\Delta} 
&=&  (2\mathcal{A}_1-\mathcal{A}_3) 
\, U^\a U^\d U^{[\b} \omega^{\g]}_{\HP\d} 
+ \mathcal{A}_3 \Big( 
\Delta^{\a\d} U^{[\b} \omega^{\g]}_{\HP\d}\nonumber\\
& & + U^\a \Delta^{\d[\b} \omega^{\g]}_{\HP\d}
+ U^\d \Delta^{\a[\b} \omega^{\g]}_{\HP\d}\Big).
\lab{eq:SDeltaGLW} 
\eeq
\end{subequations}
The thermodynamic coefficients that appear above can be expressed 
as follows (see Appendix \ref{app:spin} for details) 
\beq
\mathcal{A}_1 &=& \frac{\mathbf{\mathfrak{s}}^2}{9}\left[ \left(\frac{\partial \mathcal{N}}{\partial \xi}\right)_{\beta} - \frac{2}{m^2} \left( \frac{\partial \mathcal{E}}{\partial \beta}
\right)_{\xi}
\right],\nn\\
\mathcal{A}_3 &=& \frac{2\mathbf{\mathfrak{s}}^2}{9}\left[ \left(\frac{\partial \mathcal{N}}{\partial \xi}\right)_{\beta} + \frac{1}{m^2} \left( \frac{\partial \mathcal{E}}{\partial \beta}
\right)_{\xi}
\right],
\label{eq:A1&A3}
\eeq
where we used general expressions for $\mathcal{A}_1$ and $\mathcal{A}_3$ which are independent of the underlying statistics of the kinetic model.\footnote{See, e.g., Refs.~\cite{Florkowski:2018ahw,Bhadury:2020cop,Florkowski:2018fap,Florkowski:2021wvk} for the corresponding expressions for the MJ statistics of an ideal gas, which we summarize in Eq.~\ref{eq:A1&A3_MJ}. The case of the FD statistics is discussed in Appendix~\ref{app:FD}.}
In the above formula, $\xi = \mu / T$ is the ratio of the 
chemical potential to the temperature,
$\beta$ is the inverse of the temperature, while
$\mathbf{\mathfrak{s}}^2 = s(s+1)$ is the magnitude 
of spin angular momentum, which is equal to $3/4$ for
spin-$\frac{1}{2}$ particles~\cite{Florkowski:2018fap}.
For future convenience, we also introduce $z = m /T$
representing the ratio of particle mass $m$ and the temperature.

The (antisymmetric) spin polarization tensor $\omega_{\mu\nu}$ 
can be decomposed as~\cite{Florkowski:2017ruc}
\beq
\omega_{\mu\nu} &=& \kappa_\mu U_\nu - \kappa_\nu U_\mu + \epsilon_{\mu\nu\a\b} U^\a \omega^{\b}. \lab{spinpol1}
\eeq
where $\kappa_\mu$ and $\omega_\mu$ together form six independent 
components~\cite{Florkowski:2018fap,Florkowski:2018ahw}. 
These four-vectors are orthogonal to $U^{\mu}$ by construction,
$\kappa_\mu~U^\mu = \omega_\mu~U^\mu = 0$, such that
\cite{Florkowski:2018fap,Florkowski:2018ahw}
\beq
\kappa_\mu= \omega_{\mu\a} U^\a, \quad \omega_\mu = \half \epsilon_{\mu\a\b\gamma} \omega^{\a\b} U^\gamma. \lab{eq:kappaomega}
\eeq
In the fluid rest frame, $\kappa^{\mu}$ and $\omega^{\mu}$ 
reduce to
\beq
\kappa^\mu= \left(0, C_{\boldsymbol{\kappa}}\right), \quad \omega^\mu = \left(0, C_{\boldsymbol{\omega}}\right)\,, \lab{eq:kappaomega1}
\eeq
where $C_{\boldsymbol{\kappa}} =(C_{\kappa X}, C_{\kappa Y}, C_{\kappa Z})$ and $C_{\boldsymbol{\omega}} =(C_{\omega X}, C_{\omega Y}, C_{\omega Z})$ are the spin polarization components~\cite{Florkowski:2018fap,Florkowski:2018ahw}.
%
\section{Wave analysis}
\label{sec:wave_analysis}
\subsection{Dispersion relation for the spin modes} 
\label{sec:wave_analysis:cspin}
Let us now consider the propagation of infinitesimal excitations 
in a fluid with spin degrees of freedom. Since the conservation equations 
(\ref{Ncon}) corresponding to the background fluid are independent of polarization~\cite{Florkowski:2018ahw,Florkowski:2018fap}, 
their solutions will give the well-known spectrum of sound waves 
\cite{cercignani2002relativistic,rezzolla2013relativistic,Monnai:2014qaa,Ambrus:2017keg}, which propagate with the sound speed satisfying
\begin{eqnarray}
 c_s^2 = \left(\frac{\partial \mathcal{P}}{\partial \mathcal{E}}\right)_{\mathcal{N}} + 
 \frac{\mathcal{N}}{\mathcal{E} + \mathcal{P}} 
 \left(\frac{\partial \mathcal{P}}{\partial \mathcal{N}}\right)_{\mathcal{E}}.
\end{eqnarray}

Focusing now on the excitations propagating at the level of the spin tensor \eqref{eq:SGLW}, the background fluid can be regarded as quiescent, i.e., $U^\mu = g^{t\mu}$. Treating $\omega^{\mu\nu}$ as a small quantity, which amounts to assuming that the background fluid is unpolarized, Eqs.~\eqref{Spheno} and \eqref{eq:SDeltaGLW} reduce to
\begin{eqnarray}
 S^{\alpha,\mu\nu}_{\rm ph} &=& (\mathcal{A}_1 + \mathcal{A}_3) g^{t\alpha} \omega^{\mu\nu},\nonumber\\
 S^{\alpha,\mu\nu}_{\Delta} &=& 2(\mathcal{A}_1 - 2\mathcal{A}_3)
 g^{t\alpha} g^{t[\mu} \omega^{\nu]t} \nonumber\\
 && + \mathcal{A}_3 (g^{t[\mu} \omega^{\nu] \alpha} + 
 g^{\alpha [\mu} \omega^{\nu]t} - 
 g^{t\alpha} \omega^{\mu\nu}).
 \label{eq:S_lin}
\end{eqnarray}
Considering that the system is homogeneous with respect to the 
$x$ and $y$ directions, the divergence of Eq.~\eqref{eq:S_lin} 
yields
\begin{eqnarray}
 \partial_\alpha S^{\alpha,\mu\nu}_{\rm ph} &=& (\mathcal{A}_1  + \mathcal{A}_3) \partial_t \omega^{\mu\nu},\nonumber\\
 \partial_\alpha S^{\alpha,\mu\nu}_{\Delta} &=& 
 (2\mathcal{A}_1 - 3\mathcal{A}_3) g^{t[\mu} \partial_t \omega^{\nu] t} \nonumber\\
 && + \mathcal{A}_3 (\partial^{[\mu} \omega^{\nu] t} - 
 \partial_t \omega^{\mu\nu} + g^{t[\mu} \partial_z \omega^{\nu] z}).
\end{eqnarray}
For the cases $\mu = 0, \nu = i$ and $\mu = i, \nu = j$, we find, respectively,
\begin{eqnarray}
 \partial_\alpha S^{\alpha,ti} &=& \mathcal{A}_3\left(\partial_t \omega^{ti} + \frac{1}{2} \partial_z \omega^{i z}\right), \nonumber\\ 
 \partial_\alpha S^{\alpha,ij} &=& \mathcal{A}_1 
 \partial_t \omega^{ij} + \mathcal{A}_3 \partial^{[i} \omega^{j] t}.
 \label{eq:dSaux}
\end{eqnarray}
\begin{figure}[t]
\centering
\includegraphics[width=\columnwidth]{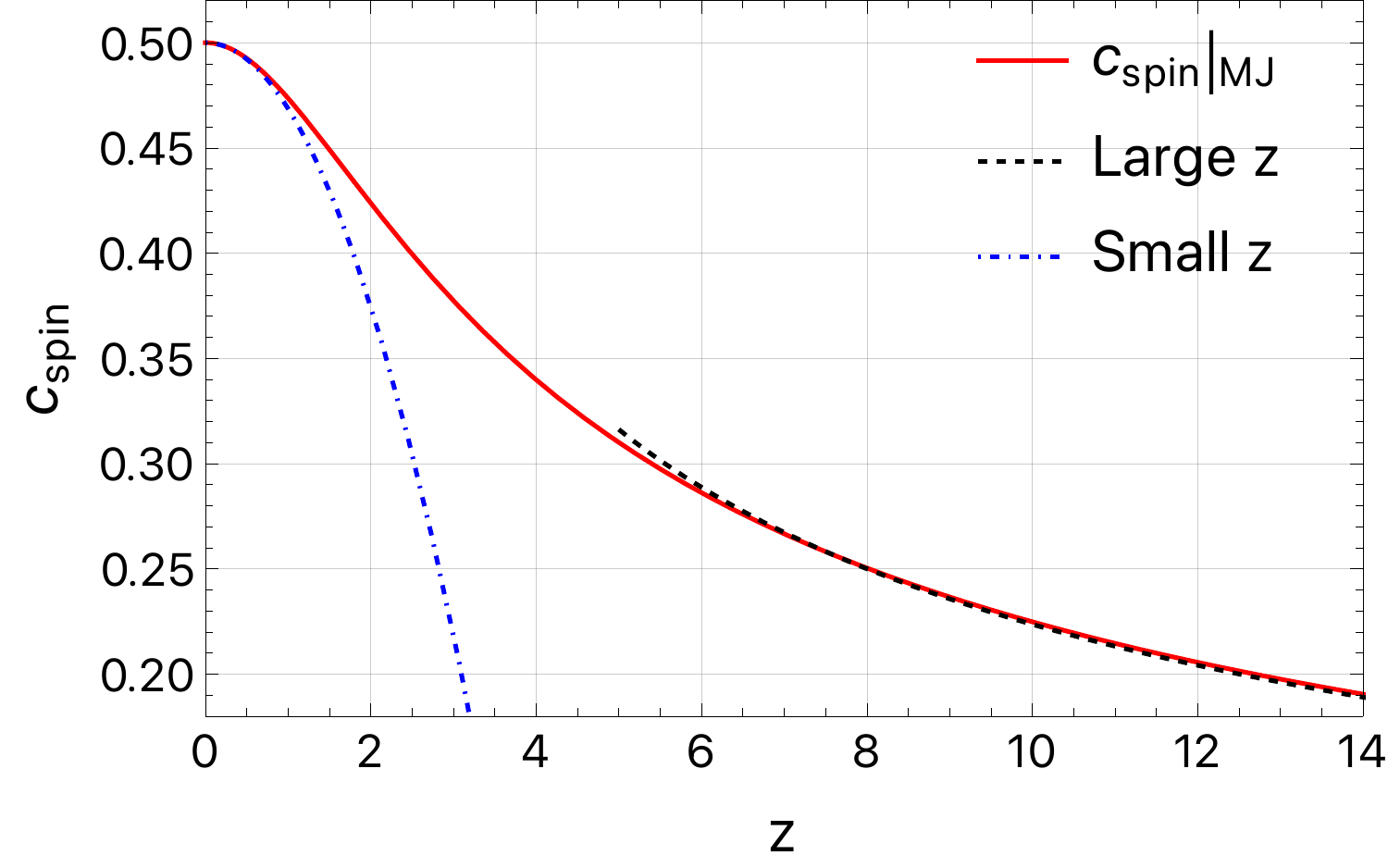}
\includegraphics[width=\columnwidth]{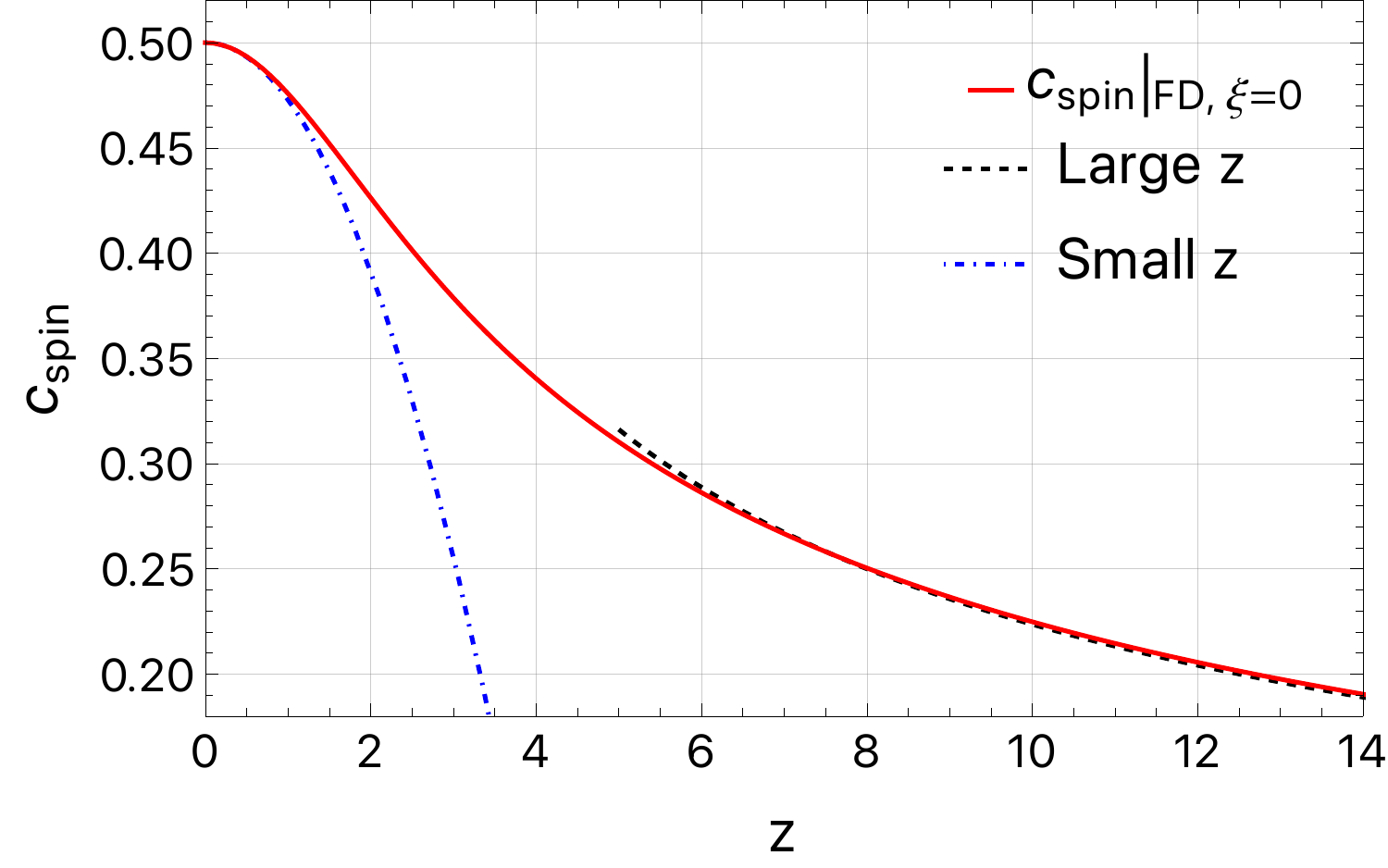}
\includegraphics[width=\columnwidth]{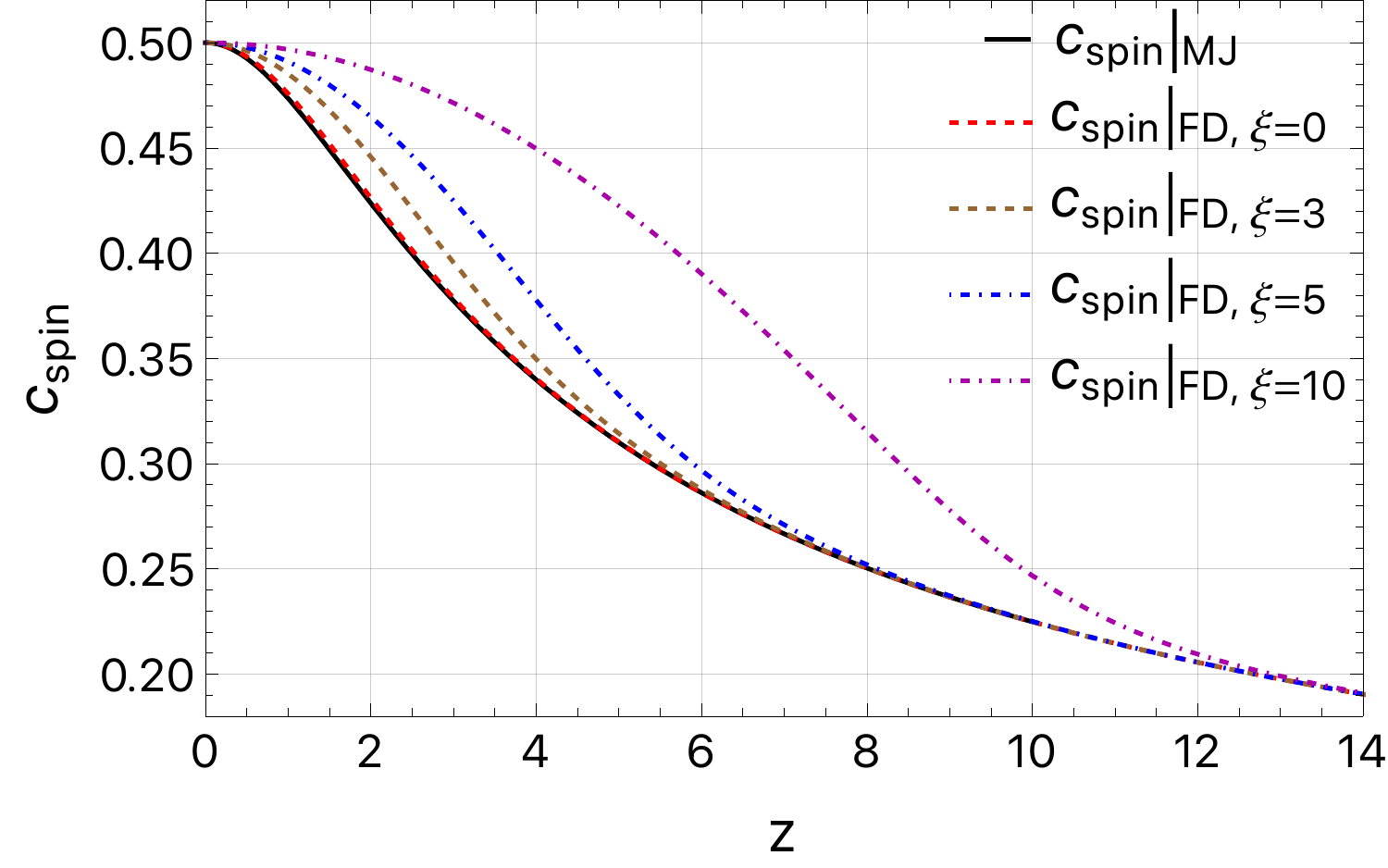}
\caption{Top, middle: the speed of the spin wave $c_{\rm spin}$ as a function of $z = m / T$ 
corresponding to the (top) MJ and (middle) FD statistics, together with the asymptotic forms for small and large $z$ given in Eqs.~\eqref{eq:cspin_smallz} and \eqref{eq:cspin_largez}, respectively.
Bottom: comparison between $c_{\rm spin}$ obtained for 
the MJ and FD statistics,
for various values of $\xi = \mu /T$ (the MJ result is independent
of $\xi$). 
The MJ curves are obtained using Eq.~\eqref{eq:cspin_Kideal}, while the FD curves are obtained using Eqs.~\eqref{eq:cspin_KFD} and \eqref{eq:cspin_FD_smallz} when $z > |\xi|$ and $z < |\xi|$, respectively.
\label{fig:cspin}
}
\end{figure}
Taking into account Eq.~\eqref{eq:kappaomega1}, the components of 
the spin polarization tensor 
$\omega^{\mu\nu}$ can be written in 
terms of the spin polarization components 
$C_{\kappa i}$ and $C_{\omega k}$
as
\begin{eqnarray}
 \omega^{ti} = -C_{\kappa i}, \qquad 
 \omega^{ij} = -\epsilon^{tijk} C_{\omega k}.
\end{eqnarray}
Demanding that
$\partial_\alpha S^{\alpha,\mu\nu} = 0$, we obtain
\begin{eqnarray}
 \partial_t C_{\kappa i} - \frac{1}{2}\epsilon^{tijz} \partial_z C_{\omega j} 
 = 0, \quad
 \partial_t C_{\omega i} - 
 \frac{\mathcal{A}_3}{2\mathcal{A}_1} 
 \epsilon^{tijz} \partial_z C_{\kappa j} = 0.
 ~~~~~~~\label{eq:eqC}
\end{eqnarray}
Due to the presence of the Levi-Civita symbol, 
$\partial_t C_{\kappa Z} = \partial_t C_{\omega Z} = 0$, such 
that the longitudinal components do not propagate.
Thus, the polarization degrees of freedom propagate only as 
transverse waves, similar to the electromagnetic waves~\cite{Jackson:1998}.
Their equation can be obtained by setting $i = x, y$ in 
Eq.~\eqref{eq:eqC}, leading to
\begin{eqnarray}
 \left(\frac{\partial^2}{\partial t^2} - c_{\rm spin}^2 \frac{\partial^2}{\partial z^2}\right) \mathcal{C} = 0,
\end{eqnarray}
where $\mathcal{C} \in \{C_{\kappa X}, C_{\kappa Y}, C_{\omega X}, C_{\omega Y}\}$ and the speed of the spin wave satisfies
\begin{eqnarray}
 c_{\rm spin}^2 = -\frac{1}{4} \frac{\mathcal{A}_3}
 {\mathcal{A}_1} = 
 \frac{1}{4} \frac{(\partial \mathcal{E} / \partial T)_\xi - 
 z^2 (\partial \mathcal{N} / \partial \xi)_T} 
 {(\partial \mathcal{E} / \partial T)_\xi + 
 \frac{z^2}{2} (\partial \mathcal{N} / \partial \xi)_T}.
 \label{eq:cspin}
\end{eqnarray}
In the ultrarelativistic limit $z \rightarrow 0$, 
we can observe that $c_{\rm spin}$ takes the value 
$1/2$ irrespective of statistics. 

The expression for $c_{\rm spin}^2$ can be written explicitly for the (ideal) MJ gas~\cite{Juttner:1911aa} 
\begin{eqnarray}
 c_{\rm spin}^2 = \frac{1}{4} \frac{K_3(z)}{K_3(z) + \frac{z}{2} K_2(z)},\label{eq:cspin_Kideal}
\end{eqnarray}
hence being independent of $\xi = \mu / T$. In the case of the FD gas~\cite{zannoni1999quantization,Dirac1926}, $c_{\rm spin}$ becomes an even function of $\xi$:\footnote{Equation~\eqref{eq:cspin_KFD} is 
valid only when $|\xi| < z$. At higher values of $|\xi|$, the formal series with respect to $\ell$ diverge and the integral representation in Eq.~\eqref{eq:cspin_FD_smallz} must be employed.}
\begin{eqnarray}
 c_{\rm spin}^2 = 
 \frac{\frac{1}{4} \sum_{\ell = 1}^\infty 
 \frac{(-1)^{\ell+1}}{\ell} \cosh(\ell\xi) K_3(\ell z)}
 {\sum_{\ell = 1}^\infty 
 \frac{(-1)^{\ell+1}}{\ell} \cosh(\ell\xi) 
 [K_3(\ell z) + \frac{\ell z}{2} K_2(\ell z)]}.
~~~ \label{eq:cspin_KFD}
\end{eqnarray}
For small values of $z$, we find
\begin{align}
 \text{MJ:}& & c_{\rm spin} =& \frac{1}{2} 
 \left[1 - \frac{z^2}{16} + O(z^4)\right],\nonumber\\
 \text{FD:}& & c_{\rm spin} =& \frac{1}{2} 
 \left[1 - \frac{15z^2}{4\pi^2} \frac{1 + \frac{3\xi^2}{\pi^2}}
 {7 + 30 \frac{\xi^2}{\pi^2} + 15 \frac{\xi^4}{\pi^4}} + O(z^4)\right].
 \label{eq:cspin_smallz}
\end{align}
In the nonrelativistic limit, when $z \gg 1$, we get
\begin{align}
 c_{\rm spin} \simeq \frac{1}{\sqrt{2z}}\,. 
\label{eq:cspin_largez}
\end{align}
The details of these calculations are provided in
Appendices~\ref{app:ideal} and \ref{app:FD}.
The above limits are validated by comparison with the exact
expressions in Eqs.~\eqref{eq:cspin_Kideal} and \eqref{eq:cspin_KFD} in Figs.~\ref{fig:cspin}. It can be observed that $c_{\rm spin}$ is a monotonically decreasing function of $z$, such that 
\begin{eqnarray}
 0 < c_{\rm spin} \le \frac{1}{2}\,,
\end{eqnarray}
where the lower limit is reached for a cold gas of massive 
particles (the nonrelativistic limit), while the upper limit
is achieved at high temperatures or for massless particles.
\subsection{Linear and circular polarization of spin waves}
\label{sec:wave_analysis:circular_pol}
%
We now construct explicit expressions for the spin wave. 
Taking as before a wave propagating along the 
$z$ direction, Eqs.~\eqref{eq:eqC} reduce to
\begin{eqnarray}
 \partial_t C_{\kappa X} - \frac{1}{2} \partial_z C_{\omega Y} &=& 0, \quad \frac{1}{2} \partial_t C_{\omega Y} - c_{\rm spin}^2 \partial_z C_{\kappa X} = 0,\nonumber\\
 \partial_t C_{\kappa Y} + \frac{1}{2} \partial_z C_{\omega X} &=& 0, \quad  \frac{1}{2} \partial_t C_{\omega X} + c_{\rm spin}^2 \partial_z C_{\kappa Y} = 0.
\end{eqnarray}
The linearly polarized solutions for the three-vectors 
$C_{\boldsymbol{\kappa}}$ and $C_{\boldsymbol{\omega}}$ 
are
\begin{eqnarray}
 C_{\boldsymbol{\kappa}} &=& C_0 {\rm Re}[e^{-i k (c_{\rm spin} t - z)}]
 (\bm{e}_1 \cos \theta + \bm{e}_2 \sin\theta), \nonumber\\
 C_{\boldsymbol{\omega}} &=& 2 c_{\rm spin} C_0 {\rm Re}[e^{-i k (c_{\rm spin} t - z)}]
 (\bm{e}_1 \sin \theta - \bm{e}_2 \cos\theta),
\end{eqnarray}
where $C_0$ is the real amplitude of the wave and $\theta$ is the 
inclination angle with respect to the $x$ axis. It can be observed that 
\begin{eqnarray}
 C_{\boldsymbol{\omega}} = 2 c_{\rm spin} \hat{\bm{n}} \times C_{\boldsymbol{\kappa}}\,,
 \label{eq:Comega_from_kappa}
\end{eqnarray}
where $\hat{\bm{n}} = \bm{e}_3$ is the direction vector of the wave.
The above equation is analogous to the 
relation $\bm{H} = c \hat{\bm{n}} \times \bm{D}$ from electromagnetism~\cite{Jackson:1998} where $c$ is the speed of light.
 
Right- and left-handed (R / L) circularly polarized waves can be constructed in the standard fashion,
\begin{eqnarray}
 C_{{\boldsymbol{\kappa}};R/L} &=& \frac{C_0}{\sqrt{2}} {\rm Re} [e^{-i k (c_{\rm spin} t - z)} (\bm{e}_1 \cos\theta \pm 
 i \bm{e}_2 \sin\theta)],
 \label{eq:circularsolns}\\
 C_{{\boldsymbol{\omega}};R/L} &=& \frac{2 c_{\rm spin} C_0}{\sqrt{2}} {\rm Re} [e^{-i k (c_{\rm spin} t - z)} (\bm{e}_1 \sin\theta \mp 
 i \bm{e}_2 \cos\theta)],\nn
\end{eqnarray}
where again Eq.~\eqref{eq:Comega_from_kappa} holds.
%
\section{Dissipative effects}
\label{sec:diss}
\begin{figure}
    \centering
    \includegraphics[width=\columnwidth]{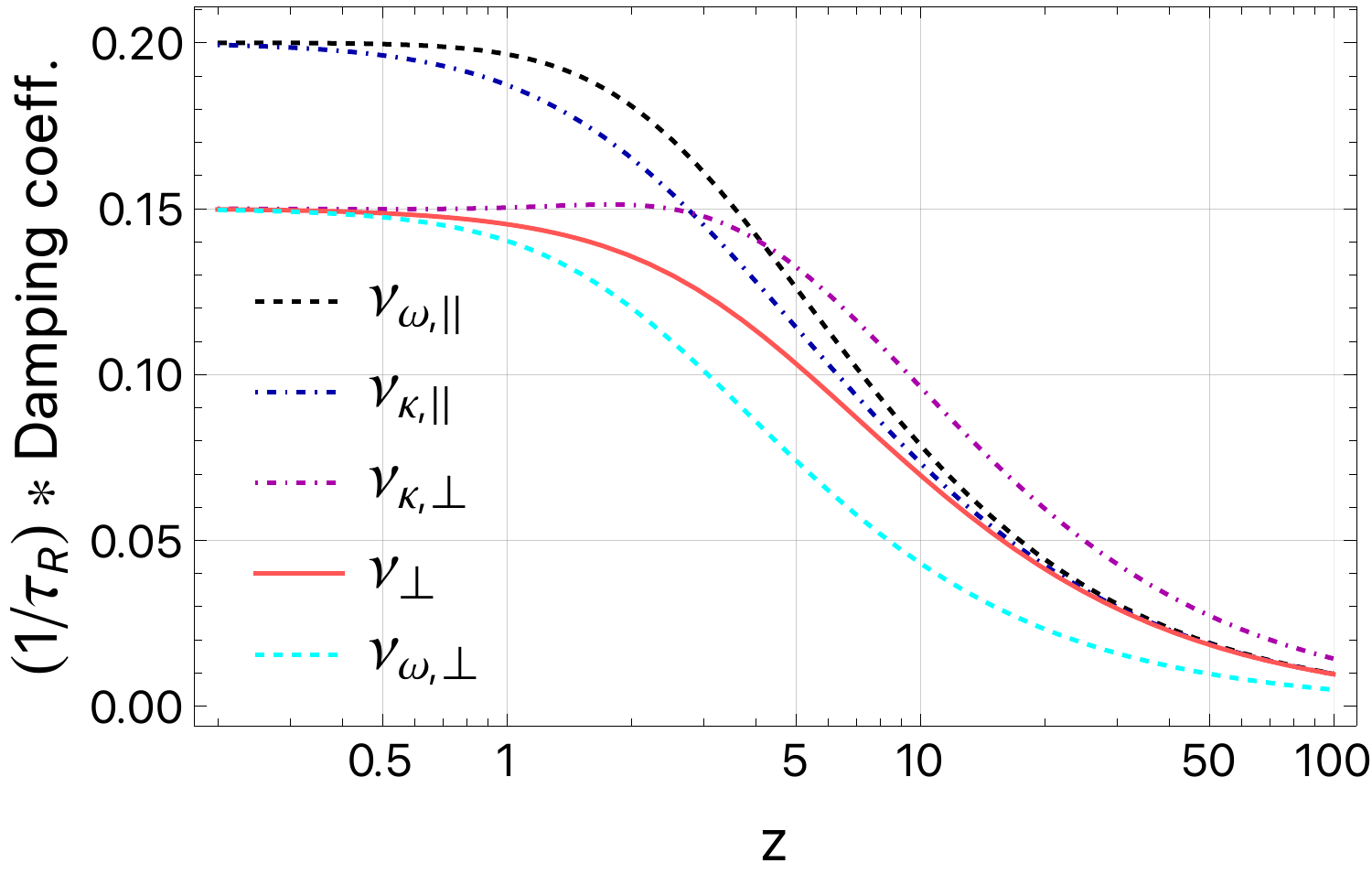}
    \caption{The $z$ dependence of the damping coefficients for the longitudinal ($\nu_{\omega,||}$, $\nu_{\kappa,||}$, starting at $0.2$) and transverse ($\nu_{\kappa,\perp}$, $\nu_\perp$ and $\nu_{\omega,\perp}$, starting at $0.15$)
    modes calculated for $\mathbf{\mathfrak{s}}^2 = 3/4$ based on Eqs.~\eqref{eq:diss_nu}, \eqref{eq:nulong}, and \eqref{eq:nuperp}.
    }
    \label{fig:damp}
\end{figure}
In this section we consider the effects of dissipation on the propagation of spin modes. In performing this analysis, we rely on the analysis of dissipative effects presented in 
Ref.~\cite{Bhadury:2020cop}. Since Ref.~\cite{Bhadury:2020cop} employs the MJ statistics of the ideal gas, we restrict the discussion in this section to this particular case. 

In the context of the relaxation time approximation, the dissipative corrections to $T^{\mu\nu}$ and $N^\mu$
turn out to be independent of the spin tensor. The correction to the spin term 
due to dissipation can be written as follows
\begin{multline}
 \delta S^{\lambda,\mu\nu}= \tau_R (B^{\lambda,\mu\nu}_\Pi \theta + 
 B^{\kappa\lambda,\mu\nu}_n \nabla_\kappa \xi + 
 B^{\kappa\delta\lambda,\mu\nu}_\pi \sigma_{\kappa\delta} \\
 + B^{\eta\beta\gamma\lambda,\mu\nu}_\Sigma 
 \nabla_\eta \omega_{\beta\gamma}),
 \label{eq:Bhadury}
\end{multline}
where $\tau_R$ is the relaxation time and $\nabla_\mu = \Delta_\mu{}^\nu \partial_\nu = \partial_\mu - U_\mu U^\nu \partial_\nu.$
As in the previous sections, 
we consider small perturbations on top of a quiescent, unpolarized background state in thermal equilibrium. 
In order to investigate the effect of dissipation on the propagation of small perturbations, and in particular to assert the stability of
the theory of spin hydrodynamics, we will regard the perturbations amplitudes (including the magnitude of $\delta S^{\lambda,\mu\nu} \sim \omega^{\mu\nu}$) as infinitesimal, while allowing the gradients, proportional to the wave number $k$, to be arbitrarily large, thereby retaining higher order terms with respect to $k$ and $\tau_R$. Within this framework, we can test if any instabilities emerge in the $k \rightarrow \infty$ limit (small wavelengths). We note that, due to the truncation procedure leading to Eq.~\eqref{eq:Bhadury}, we cannot expect the physics of the $\tau_R \gg 1$ and/or $k \gg 1$ regimes to be correctly recovered.

Coming back to Eq.~\eqref{eq:Bhadury}, the coefficients $B^{\lambda,\mu\nu}_\Pi$,
$B^{\kappa\lambda,\mu\nu}_n$, and $B^{\kappa\delta\lambda,\mu\nu}_\pi$ are proportional to the spin polarization tensor $\omega^{\mu\nu}$~\cite{Bhadury:2020cop}, which we assume to be of first order with respect to the perturbation amplitude (the background state is assumed to be unpolarized).
These terms are multiplied by $\theta = \partial_\mu u^\mu$, $\nabla_\kappa \xi$, and $\sigma_{\kappa\delta} = \frac{1}{2}(\nabla_\kappa u_\delta + \nabla_\delta u_\kappa) - \frac{1}{3} \theta \Delta_{\kappa\delta}$, respectively, which are already of first order with respect to the gradients of the background state. Since they are of second order with respect to the perturbation amplitude, the first three terms appearing in Eq.~\eqref{eq:Bhadury} can be safely neglected
and we focus only on the last term given by~\cite{Bhadury:2020cop}
\begin{multline}
 B^{\eta\beta\gamma\lambda,\mu\nu}_\Sigma = 
 B^{(1)}_\Sigma \Delta^{\lambda\eta} g^{\beta[\mu} g^{\nu] \gamma} + 
 B^{(2)}_\Sigma \Delta^{\lambda\eta} u^\gamma u^{[\mu} \Delta^{\nu]\beta} \\
 + B^{(3)}_\Sigma(\Delta^{\lambda\eta} \Delta^{\gamma[\mu} g^{\nu]\beta} + 
 \Delta^{\lambda\gamma} \Delta^{\eta[\mu} g^{\nu] \beta} + 
 \Delta^{\gamma\eta} \Delta^{\lambda[\mu} g^{\nu] \beta}) \\
 + B^{(4)}_\Sigma \Delta^{\gamma\eta} \Delta^{\lambda[\mu} \Delta^{\nu] \beta} + 
 B^{(5)}_\Sigma u^\gamma \Delta^{\lambda\beta} u^{[\mu} \Delta^{\nu]\eta}.
\end{multline}
The quantities $B^{(i)}_\Sigma$ are~\cite{Bhadury:2020cop}
\begin{align}
 B^{(1)}_\Sigma =& -\frac{4 \mathbf{\mathfrak{s}}^2}{3} \cosh \xi I^{(1)}_{21}, 
 \nonumber\\
  B^{(2)}_\Sigma =& -\frac{8 \mathbf{\mathfrak{s}}^2}{3m^2} \cosh \xi \left(I^{(1)}_{41} + \frac{I^{(1)}_{41} I^{(0)}_{31}}
  {m^2 I^{(0)}_{10} - 2 I^{(0)}_{31}}\right), 
  \nonumber\\
  B^{(3)}_\Sigma =& -\frac{8 \mathbf{\mathfrak{s}}^2}{3m^2} \cosh \xi I^{(1)}_{42},
  \nonumber\\
  B^{(4)}_\Sigma =& -\frac{8 \mathbf{\mathfrak{s}}^2}{3m^2} \cosh \xi 
  \frac{I^{(1)}_{41} I^{(0)}_{31}}{m^2 I^{(0)}_{10} - (I^{(0)}_{30} 
  + I^{(0)}_{31})},\nonumber\\
  B^{(5)}_\Sigma =& \frac{8 \mathbf{\mathfrak{s}}^2}{3m^2} \cosh \xi 
  \frac{I^{(1)}_{41} I^{(0)}_{31}}{m^2 I^{(0)}_{10} - 2I^{(0)}_{31}},
\end{align}
where $I^{(r)}_{nq}$ are thermodynamic integrals of the form
\begin{eqnarray}
 I^{(r)}_{nq} = \frac{1}{(2q+1)!!} 
 \int \mathrm{dP} (u\cdot p)^{n-2q-r} (\Delta_{\alpha\beta} p^\alpha p^\beta)^q e^{-\beta u \cdot p},\nonumber\\
\end{eqnarray}
and $\mathrm{dP} = d^3p / (2\pi)^3 p^0$ defines  the invariant integration measure. Since $A^\eta \nabla_\eta \omega_{\beta\gamma}$ reduces to $A^z \partial_z \omega_{\beta\gamma}$, the index $\eta$ can be safely set to $z$. Performing the splitting
\begin{eqnarray}
 \partial_\lambda \delta S^{\lambda,\mu\nu} = \tau_R \sum_i B_\Sigma^{(i)} 
 \mathfrak{T}^{(i) \mu\nu},
\end{eqnarray}
we find 
\begin{align}
 \mathfrak{T}^{(1)\mu\nu} =& -\partial_z^2 \omega^{\mu\nu}, \nonumber\\
 \mathfrak{T}^{(2)\mu\nu} =& -g^{t[\mu} \partial_z^2 \omega^{\nu]t}, \nonumber\\
 \mathfrak{T}^{(3)\mu\nu} =& \partial_z^2 \omega^{\mu\nu} + 
 g^{t[\mu} \partial^2_z \omega^{\nu]t} + 
 2g^{z[\mu} \partial^2_z \omega^{\nu] z},\nonumber\\\
 \mathfrak{T}^{(4)\mu\nu} =& g^{z[\mu} \partial^2_z \omega^{\nu]z} - g^{z[\mu} g^{\nu]t} \partial^2_z \omega^{tz}, \nonumber\\
 \mathfrak{T}^{(5)\mu\nu} =& g^{z[\mu} g^{\nu]t} \partial^2_z \omega^{tz}.
\end{align}
Grouping all terms together gives
\begin{multline}
 \frac{1}{\tau_R} \partial_\lambda \delta S^{\lambda,\mu\nu} = 
 -\LR B^{(1)}_\Sigma - B^{(3)}_\Sigma\RR \partial_z^2 \omega^{\mu\nu} \\
 - \LR B^{(2)}_\Sigma - B^{(3)}_\Sigma\RR g^{t[\mu} \partial^2_z \omega^{\nu] t}
 + 2B^{(3)}_\Sigma g^{z[\mu} \partial^2_z \omega^{\nu] z} \\
 -\LR B^{(4)}_\Sigma - B^{(5)}_\Sigma\RR 
 g^{z[\mu} g^{\nu] t} \partial^2_z \omega^{tz} + 
 B^{(4)}_\Sigma g^{z[\mu} \partial^2_z \omega^{\nu] z}.
\end{multline}
Noting that $\omega^{ti} = -C_{\kappa i}$ and $\omega^{ij} = -\epsilon^{0ijk} C_{\omega k}$, we have:
\begin{align}
 \partial_\lambda \delta S^{\lambda, t x} =& \nu_{\kappa,\perp} \mathcal{A}_3 
 \partial^2_z C_{\kappa X}, &
 \partial_\lambda \delta S^{\lambda, yz} =& 
 \nu_{\omega,\perp} \mathcal{A}_1 \partial^2_z C_{\omega X}, \nonumber\\
 \partial_\lambda \delta S^{\lambda, t y} =& \nu_{\kappa,\perp} \mathcal{A}_3 
 \partial^2_z C_{\kappa Y}, & 
 \partial_\lambda \delta S^{\lambda, zx} =& 
 \nu_{\omega,\perp} \mathcal{A}_1 \partial^2_z C_{\omega Y},
 \nonumber\\
 \partial_\lambda \delta S^{\lambda, tz} =& \nu_{\kappa, ||} \mathcal{A}_3
 \partial^2_z C_{\kappa Z}, & 
 \partial_\lambda \delta S^{\lambda, xy} =& 
 \nu_{\omega,||} \mathcal{A}_1 \partial^2_z C_{\omega Z}.
 \label{eq:diss_eqs_aux}
\end{align}
In Eq.~\eqref{eq:diss_eqs_aux} we identified the longitudinal ($\nu_{\kappa,||}$, $\nu_{\omega,||}$) and transverse ($\nu_{\kappa,\perp}$, $\nu_{\omega,\perp}$) 
kinematic viscosities,
\begin{align}
 \nu_{\kappa, ||} =& \frac{\tau_R}{\mathcal{A}_3} 
 \left(B^{(1)}_\Sigma - \frac{1}{2} B^{(2)}_\Sigma - \frac{3}{2} 
 B^{(3)}_\Sigma - \frac{1}{2} B^{(5)}_\Sigma\right) = \frac{\tau_R}{\mathcal{A}_3} B^{(3)}_\Sigma,\nonumber\\
 \nu_{\omega,||} =& \frac{\tau_R}{\mathcal{A}_1} \left(B^{(1)}_\Sigma - B^{(3)}_\Sigma\right),\nonumber\\
 \nu_{\kappa,\perp} =& \frac{\tau_R}{\mathcal{A}_3} \left(B^{(1)}_\Sigma - \frac{1}{2} B^{(2)}_\Sigma - 
 \frac{1}{2} B^{(3)}_\Sigma\right),\nonumber\\
 \nu_{\omega,\perp} =& \frac{\tau_R}{\mathcal{A}_1} 
 \left(B^{(1)}_\Sigma - 2 B^{(3)}_\Sigma - 
 \frac{1}{2} B^{(4)}_\Sigma \right).
 \label{eq:diss_nu}
\end{align}
The expression for $\nu_{\kappa,||}$ follows after applying
the recurrence relation $I^{(r)}_{nq} = \frac{1}{2q+1}
\LR m^2 I^{(r)}_{n-2,q-1} - I^{(r)}_{n,q-1}\RR$ \cite{Bhadury:2020cop}
to $I^{(1)}_{42}$.

Combining the above results with Eq.~\eqref{eq:dSaux}, one can find that $C_{\kappa Z}$ and $C_{\omega Z}$ 
exhibit exponential decay,
\begin{eqnarray}
 \partial_t C_{\kappa Z} - \nu_{\kappa,||} \partial^2_z C_{\kappa Z} = 0,\quad
 \partial_t C_{\omega Z} - \nu_{\omega,||} \partial^2_z C_{\omega Z} = 0.~~~~
\end{eqnarray}
Setting now $C_{\kappa/\omega; Z}\!\! \sim \!\! e^{-i \omega t + i k z} \widetilde{C}_{\kappa/\omega; Z}$, 
where $\widetilde{C}_{\kappa/\omega; Z}$ is a constant, we find $\omega = -i k^2 \nu_{\kappa/\omega, ||}$ with
\begin{eqnarray}
 \nu_{\kappa,||} &=& \frac{4\mathbf{\mathfrak{s}}^2 \tau_R}{45 G(z)} 
 [-5z + G(z)(3 + z^2) - z^2 {\rm Gi}(z)] \nonumber\\
 &\simeq& \frac{4 \mathbf{\mathfrak{s}}^2 \tau_R}{15}
 \left[1 - \frac{z^2}{12} + O(z^4)\right],\nonumber\\
 \nu_{\omega,||} &=& \frac{4\mathbf{\mathfrak{s}}^2 \tau_R}{15(2G(z) + z)} 
 [5z + G(z)(2 - z^2) + z^2 {\rm Gi}(z)] \nonumber\\
 &\simeq& \frac{4 \mathbf{\mathfrak{s}}^2 \tau_R}{15}
 \left[1 - \frac{z^4}{16} + O(z^5)\right],
 \label{eq:nulong}
\end{eqnarray}
where $G(z) = K_3(z) / K_2(z)$ and ${\rm Gi}(z) = {\rm Ki}_1(z) / K_2(z)$. The above expressions are represented as functions of $z$ in Fig.~\ref{fig:damp}.

Performing now the Fourier decomposition 
$\mathcal{C} = \widetilde{\mathcal{C}} e^{-i \omega t + i k z}$ of the transverse modes, we find
\beq
 \begin{pmatrix}
  \omega + i k^2 \nu_{\kappa, \perp} & -k/2\\
  \frac{k \mathcal{A}_3}{2 \mathcal{A}_1} & \omega + i k^2 \nu_{\omega,\perp} 
 \end{pmatrix}
 \begin{pmatrix}
 \widetilde{C}_{\kappa X} \\
 \widetilde{C}_{\omega Y} 
 \end{pmatrix} &=& 0,\nonumber\\
 \begin{pmatrix}
  \omega + i k^2 \nu_{\kappa,\perp} & k/2 \\
  - \frac{k \mathcal{A}_3}{2 \mathcal{A}_1} & 
  \omega + i k^2 \nu_{\omega,\perp}
 \end{pmatrix}
 \begin{pmatrix}
 \widetilde{C}_{\kappa Y} \\
 \widetilde{C}_{\omega X}
 \end{pmatrix} &=& 0.
\eeq
The dispersion relation implied by the above system is
\begin{align}
 \omega_\pm =& -i k^2 \nu_\perp \pm k c_{\rm spin}\,, \qquad 
 \nu_\perp = \frac{\nu_{\kappa,\perp} + \nu_{\omega,\perp}}{2}\,,
 \nonumber\\
 c_{\rm spin}^2 =& -\frac{\mathcal{A}_3}{4\mathcal{A}_1} - \frac{k^2}{4} \left(\nu_{\kappa,\perp} - \nu_{\omega,\perp}\right)^2.
 \label{eq:diss_disp}
\end{align}
The damping coefficient $\nu_\perp$ is just the average of the
damping coefficients found separately for the $\kappa$ and 
$\omega$ sectors,
\begin{align}
 \nu_\perp =& \frac{2 \mathbf{\mathfrak{s}}^2 \tau_R [3G(z) + 2z]}{45G(z)[2G(z) + z]} 
 [- 5z + G(z)(3 + z^2) - z^2 {\rm Gi}(z)]\nonumber\\
 \simeq& \frac{\mathbf{\mathfrak{s}}^2 \tau_R}{5} 
 \left[1 - \frac{z^2}{24} + O(z^4)\right].
 \label{eq:nuperp}
\end{align}
The above expression is represented as a function of $z$ in Fig.~\ref{fig:damp}.

The speed of the spin wave receives a dissipative correction of negative sign,
which can be estimated by writing $c_{\rm spin}^2 = c^2_{\rm spin; 0}
(1 - \delta c^2_{\rm spin})$,
where $c_{\rm spin;0}^2 = -\mathcal{A}_3 / 4 \mathcal{A}_1 > 0$ and 
\begin{equation}
 \delta c^2_{\rm spin} = \frac{k^2 (\nu_{\kappa,\perp} - \nu_{\omega,\perp})^2}{-\mathcal{A}_3 / \mathcal{A}_1}
 \simeq \frac{k^2 \mathbf{\mathfrak{s}}^4 \tau_R^2}{8100} z^4[1 + O(z^6)].
\end{equation}
The above correction is heavily suppressed 
at small values of $z$. At finite $z$, the 
wave number can become large enough to render 
$c_{\rm spin}^2$ negative. This happens when 
$k$ exceeds the threshold value given by
\begin{eqnarray}
k_{\rm th} = \frac{2 c_{\rm spin;0}}{|\nu_{\kappa,\perp} - \nu_{\omega,\perp}|}.
\end{eqnarray}
When $k > k_{\rm th}$, $c_{\rm spin}$ becomes imaginary and the wave no longer propagates. This is reminiscent of similar effects occurring in first-order hydrodynamics for spinless systems. One example is the case of sound modes in ultrarelativistic fluids, where $\tau_R k_{\rm th} = \frac{5\eta}{4\mathcal{P}} k_{\rm th} = 15/2$ \cite{Ambrus:2017keg}. Considering now the regime when $k \gg k_{\rm th}$, Eq.~\eqref{eq:diss_disp} shows that the modes remain stable provided 
\begin{eqnarray}
 \nu_\perp - \frac{1}{2} |\nu_{\kappa,\perp} - \nu_{\omega,\perp} |  = 
 {\rm min}(\nu_{\kappa,\perp}, \nu_{\omega,\perp}) > 0.
\end{eqnarray}
The above inequality holds true within the framework studied here. We show this in the regime of small $z$, when 
\begin{eqnarray}
 \nu_{\kappa,\perp} &\simeq& \frac{\mathfrak{s}^2 \tau_R}{5}\left[1 - \frac{z^2}{72} + O(z^4)
 \right],\nn\\
 \nu_{\omega,\perp} &\simeq& \frac{\mathfrak{s}^2 \tau_R}{5}\left[1 - \frac{5z^2}{72} + O(z^4)
 \right],
\end{eqnarray}
while $\tau_R k_{\rm th} \simeq 18 / (5 z^2 \mathfrak{s}^2)$.
Figure~\ref{fig:damp} confirms that both $\nu_{\kappa,\perp}$ and $\nu_{\omega,\perp}$ remain positive at large $z$, thus the theory is stable under linear perturbations.

Let us now consider the impact of dissipation on the propagation of the spin waves in the context of heavy-ion collisions. For simplicity, let us focus on the $z \ll 1$ case, when the shear viscosity, $\eta$, can be related to the relaxation time via $\eta = \frac{4}{5} \tau_R \mathcal{P}$ \cite{cercignani2002relativistic,Ambrus:2017keg}.
Assuming that the ratio $\eta / \mathcal{S}$ is constant, 
where $\mathcal{S} = (\mathcal{E} + \mathcal{P} - \mu \mathcal{N}) / T \simeq 4\mathcal{P} / T$ is the entropy density (we considered also $|\xi| \ll 1$), we have 
\begin{eqnarray}
 \tau_R \simeq \frac{5}{4\pi^2 T} \times (4\pi \eta /\mathcal{S}).
\end{eqnarray}
Setting now $\mathfrak{s}^2 = 3/4$, the damping time 
$t_{\rm damp; \perp} = 1 / k^2 \nu_\perp$ can be estimated as
\begin{align}
 t_{\rm damp; \perp} \simeq& \frac{4\lambda^2 T/3}{4\pi \eta / \mathcal{S}} \nonumber\\
 =& \left(\frac{\lambda}{1\ {\rm fm}}\right)^2 
 \left(\frac{T}{600\ {\rm MeV}}\right) \times 
 \frac{4\ {\rm fm} / c}{4\pi \eta / \mathcal{S}},
\end{align}
where $\lambda = 2\pi / k$ is the wavelength. Thus the lifetime of spin waves is of the same order of magnitude as 
the lifetime of the QGP fireball.
%
\section{Conclusions}
\label{sec:conclusion}
%
In this work we have studied the wave spectrum of the theory of spin hydrodynamics based on the GLW pseudogauge. As an antisymmetric tensor of rank two, the spin chemical potential $\omega^{\mu\nu}$ has six independent degrees of freedom, which can be divided into three electric and three magnetic ones. 
Our analysis has revealed the transverse nature of the spin wave. In the limiting case of the ideal fluid, the longitudinal magnetic and electric components do not propagate, while the transverse ones oscillate, leading to the linearly or circularly polarized waves known from the theory of electromagnetism.

The speed of the spin wave, $c_{\rm spin}$, generally depends on the parameters of the medium (temperature $T$, chemical potential $\mu$) and on the properties of the particles (particle mass $m$ or the statistics obeyed by the particles). A generic feature of the modes is that in the ultrarelativistic limit ($z = m/T \ll 1$), $c_{\rm spin} \simeq 1/2$, a property that is independent of the statistics. 
In the case of the ideal MJ gas, $c_{\rm spin}$ becomes independent of $\xi = \mu /T$. In the case of FD statistics, we found that the chemical potential enhances $c_{\rm spin}$ and maintains the ultrarelativistic threshold for small values of $z / \xi = m / \mu$. 
At the other end of the spectrum, when $z \gg 1$, we find the leading-order behavior $c_{\rm spin} \sim 1/\sqrt{2z}$, again independent of the statistics.

Finally, we have studied the effects of dissipation on the spin waves. At the level of first-order spin hydrodynamics, the transverse components are all damped via the same coefficient $\nu_\perp$. The longitudinal components $C_{\kappa Z}$ and $C_{\omega Z}$ decay exponentially with different coefficients, $\nu_{\kappa,||}$ and $\nu_{\omega,||}$. The speed $c_{\rm spin}$ receives a viscous correction which becomes dominant at large wave numbers $k$. Above the threshold 
$\tau_R k_{\rm th} \simeq 18 / (5z^2 \mathfrak{s}^2)$, 
$c_{\rm spin}$ becomes imaginary and the wave no longer propagates.

The approach considered in this paper, based on the spin polarization tensor $\omega^{\alpha\beta}$, does not account for anomalous transport phenomena.
The addition of vortical terms in $N^\alpha$ and $T^{\alpha\beta}$ modifies the wave spectrum corresponding to the fluid sector, giving rise to a rich spectrum of excitations, such as the chiral magnetic wave, chiral vortical wave, chiral heat wave, or helical vortical wave~\cite{Chernodub:2015gxa,Kalaydzhyan:2016dyr,Ambrus:2019khr} . An investigation of the interplay between anomalous transport effects and the dynamics of the spin polarization tensor represents an intriguing avenue for future research.
%
\begin{acknowledgments}
%
We thank D. H. Rischke, M. Shokri, and P. Aasha
for fruitful discussions and A. Palermo for the critical reading of the manuscript.
V.E.A. gratefully acknowledges the support of the
Alexander von Humboldt Foundation through a Research
Fellowship for postdoctoral researchers, as well as
the kind hospitality of the Institute of Nuclear Physics Polish Academy of Sciences, Krakow where this work was initiated.
R.S. acknowledges the support of Polish NAWA PROM Program No.: PROM PPI/PRO/2019/1/00016/U/001 and the hospitality of the Institute for Theoretical Physics, Goethe University Germany where this work was finalized.
This research was also supported in part by the Polish National Science Centre Grants No. 2016/23/B/ST2/00717 and No. 2018/30/E/ST2/00432.
\end{acknowledgments}
\appendix 
\section{Spin tensor decomposition for general statistics}
\label{app:spin}
%
We start with the equilibrium phase-space distribution function which is constructed after the identification of the collisional invariants for MJ statistics~\cite{Florkowski:2018fap,Bhadury:2020puc}
\begin{equation}
f^\pm_{\rm eq} =  \exp\left[-\beta p \cdot U \pm\xi\right] 
\exp\left[\frac{1}{2} \omega_{\mu\nu} s^{\mu\nu} \right],
\label{eq:feqxps}
\end{equation}
where $s^{\alpha\beta} = \frac{1}{m} \epsilon^{\alpha\beta\mu\nu} p_\mu s_\nu$ is the internal angular momentum
and $s^\mu$ is the spin four-vector~\cite{Mathisson:1937zz,Itzykson:1980rh}.

The spin tensor (\ref{eq:SGLW}) is derived through the moments of (\ref{eq:feqxps}) as~\cite{Florkowski:2018fap}
\beq
S^{\lambda, \mu\nu} &=& \int  \mathrm{dP}~\mathrm{dS} \, \, p^\lambda \, s^{\mu \nu} 
\left[f^+_{\rm eq} + f^-_{\rm eq} \right] \nn \\
&=& 2 \cosh \xi \int  \mathrm{dP} \, p^\lambda \exp\LB - \beta p \cdot U \RB \nonumber\\
&& \hspace{1cm} \times
\int  \mathrm{dS} \, s^{\mu \nu} \, \exp\LB \frac{1}{2}  \omega_{\alpha \beta} s^{\alpha\beta} \RB, \label{eq:Seq-sp01} 
\eeq
where  $\mathrm{dS} = \frac{m}{\pi \mathbf{\mathfrak{s}}} d^4s\,
\delta(s \cdot s + \mathbf{\mathfrak{s}}^2) \delta(p \cdot s)$ is the invariant spin measure
\cite{Florkowski:2018fap}.
To the leading order in $\omega_{\alpha\beta}$ the second integral is expressed as
\bea
&&\hspace{-0.cm}  \int  \mathrm{dS} \, s^{\mu \nu}  \, \exp\LB  \frac{1}{2}  \omega_{\alpha \beta} s^{\alpha\beta}\RB \simeq
\int  \mathrm{dS} \, s^{\mu \nu}  \, \left(1 +   \frac{1}{2}  \omega_{\alpha \beta} s^{\alpha\beta}\right)
\nn \\
&&\hspace{1.5cm}=\frac{2 \mathfrak{s}^2}{3 m^2} \LB m^2 \omnU + 2 p^\alpha p^{[\mu} \omega^{\nu ]}_{\HP\alpha} \RB,
\label{eq:spinint3}
\eea
where the integral in the spin space was performed using the following relations~\cite{Florkowski:2018fap}:
\begin{eqnarray}
 \int \mathrm{dS} &=& 2, \qquad 
 \int \mathrm{dS}\, s^\mu = 0,\nonumber\\
 \int \mathrm{dS} \, s^\mu s^\nu &=&
 \frac{2\mathfrak{s}^2}{3m^2}(p^\mu p^\nu - 
 m^2 g^{\mu\nu}),
\end{eqnarray}
leading to $\int \mathrm{dS}\, s^{\mu\nu} = 0$ and
\begin{eqnarray}
 \int \mathrm{dS}\, s^{\mu\nu} s^{\alpha\beta} 
 = \frac{4\mathfrak{s}^2}{3m^2} 
 (m^2 g^{\mu[\alpha}g^{\beta]\nu} + 
 2 p^{[\alpha} g^{\beta] [\mu} p^{\nu]}).
\end{eqnarray}
Using (\ref{eq:spinint3}) in (\ref{eq:Seq-sp01}) we have~\cite{Florkowski:2018fap}
\bea
S^{\lambda, \mu\nu} 
&=& \frac{4 {{\mathfrak{s}}}^2}{3 m^2} \cosh \xi \!\!\int \!\! \mathrm{dP} \, p^\lambda \, e^{ - p \cdot \beta }
\LB m^2 \omnU \!+\! 2 p^\alpha p^{[\mu} \omega^{\nu ]}_{\HP\alpha} \RB,
~~~~~~~\label{eq:Seq-sp1}
\eea
which is the spin tensor for the MJ statistics~(\ref{eq:SGLW}).

Now we extend the distribution function (\ref{eq:feqxps}) to general statistics, where $f_{\rm eq}^{\sigma} \equiv f_{\rm eq}^{\sigma}(y_\sigma)$, and
\begin{eqnarray}
 y_\sigma &=& y_{\sigma;0} + 
 y_{\rm spin}, \quad
 y_{\sigma;0} = \beta p \cdot U - \sigma \xi,\nn\\
 && \qquad y_{\rm spin} = 
 -\frac{1}{2} \omega^{\mu\nu} s_{\mu\nu}.
\end{eqnarray}
We consider $y_{\rm spin} \ll y_{\sigma;0}$, such that
\begin{eqnarray}
 f_{\rm eq}^\sigma(y_\sigma) = 
 f_{\rm eq}^\sigma(y_{\sigma;0}) + 
 f^{\sigma \prime}_{\rm eq}(y_{\sigma;0}) 
 y_{\rm spin} + \dots,
\end{eqnarray}
where $\sigma = +1$ and $-1$ for particles and antiparticles, respectively.
The derivative of the distribution function is 
evaluated at vanishing spin chemical potential, such that
\begin{eqnarray}
 f^{\prime\sigma}_{\rm eq}(y_{\sigma;0}) =  
 -\sigma \left(\frac{\partial f^\sigma_{\rm eq}}{\partial \xi}\right)_\beta
 = \frac{1}{p \cdot U}
 \left(\frac{\partial f^\sigma_{\rm eq}}{\partial \beta}\right)_\xi.
\end{eqnarray}
Therefore, Eq.~\eqref{eq:Seq-sp01} can be written for the general statistics as
\bea
S^{\lambda, \mu\nu}_{\rm eq} 
&=& -\frac{2 {{\mathfrak{s}}}^2}{3 m^2} \sum_{\sigma = \pm}\,\, \!\!\int \!\! \mathrm{dP} \, p^\lambda f^{\prime\sigma}_{\rm eq}
\left(m^2 \omnU \!+\! 2 p^\alpha p^{[\mu} \omega^{\nu ]}_{\HP\alpha}\right).\nn\\
~~~~\label{eq:Seq-sp11}
\eea
The integral of $p^\lambda f^{\sigma \prime}_{\rm eq}$ can be written in terms of the number density via
\begin{eqnarray}
 2\sum_{\sigma = \pm} \int \mathrm{dP}\, p^\lambda 
 f^{\sigma\prime}_{\rm eq} = 
 a_1 U^\lambda, \quad 
 a_1 = - \left(\frac{\partial \mathcal{N}}{\partial \xi}\right)_{\beta},
 \label{eq:Seq_aux_p}
\end{eqnarray}
where the factor of $2$ accounts for the spin degeneracy ($\int \mathrm{dS} = 2$) and 
we used the perfect fluid form $N^\alpha = \mathcal{N} U^\alpha$ for the charge current.
The integral involving $p^\lambda p^\mu p^\alpha f^{\prime\sigma}_{\rm eq}$ allows to perform
the tensor decomposition
\begin{multline}
2\sum_{\sigma = \pm} \int \mathrm{dP} 
f^{\sigma\prime}_{\rm eq}\, p^\lambda p^\alpha p^\mu\\ 
= a_2 \, U^\lambda U^\alpha U^\mu + 
b_2 (U^\lambda \Delta^{\alpha\mu} + U^\alpha \Delta^{\lambda\mu} 
+ U^\mu \Delta^{\lambda\alpha}),
\label{eq:Seq_aux_ppp}
\end{multline}
where the coefficients $a_2$ and $b_2$ can be obtained by
contracting the above expression with 
$U_\lambda U_\alpha U_\mu$ and $U_\lambda g_{\alpha \mu}$, 
respectively:
\beq
a_2 = \left(\frac{\partial \mathcal{E}}{\partial \mathcal{\beta}} \right)_\xi, \qquad
a_2 + 3b_2 = -m^2\left(\frac{\partial \mathcal{N}}{\partial \xi} \right)_{\beta}.
\label{eq:Seq_aux_ppp_ab}
\eeq
Substituting the above results in Eq.~\eqref{eq:Seq-sp11} and 
comparing with Eq.~\eqref{eq:S} shows that
$a_1 = -\frac{3}{\mathfrak{s}^2} (\mathcal{A}_1 + \mathcal{A}_3)$,
$a_2 = \frac{3m^2}{2\mathfrak{s}^2}(\mathcal{A}_3 - 2 \mathcal{A}_1)$, and 
$b_2 = -\frac{3m^2}{2\mathfrak{s}^2} \mathcal{A}_3$, 
where the coefficients $\mathcal{A}_1$ and $\mathcal{A}_3$ are 
given in Eq.~\eqref{eq:A1&A3} and are compatible 
with Eq.~\eqref{eq:Seq_aux_ppp_ab}.
%
\section{Ideal gas}
\label{app:ideal}
%

The ideal gas is modeled using the MJ
distribution,
\begin{eqnarray}
 f^\sigma_{\rm eq} = e^{-\beta U \cdot p + \sigma \xi},
\end{eqnarray}
where $\beta$ is the inverse of the temperature and $\xi = \mu \beta$.
The charge current $N^\mu$ and energy-momentum tensor $T^{\mu\nu}$ can be obtained as
\begin{eqnarray}
 N^\mu &=& 2 \sum_{\sigma} \sigma \int \mathrm{dP} \, p^\mu f^{\sigma}_{\rm eq}, \nonumber\\
 T^{\mu\nu} &=& 2 \sum_{\sigma} \int \mathrm{dP} \, p^\mu  p^\nu f^{\sigma}_{\rm eq},
\end{eqnarray}
where the factor of 2 accounts for the spin degeneracy. The integrals yield the perfect fluid form 
in Eq.~\eqref{Nmu}, where ${\cal N}$, ${\cal E}$, 
and ${\cal P}$ are given by
\cite{Florkowski:1321594,Florkowski:2017ruc,Florkowski:1321594}
\ba
{\cal N} &=& 4 \sinh(\xi) {\cal N}_{(0)} ,
\quad
\begin{pmatrix}
 {\cal E}  \\ {\cal P} 
\end{pmatrix}
= 4 \cosh(\xi) 
\begin{pmatrix}
{\cal E}_{(0)}\\
{\cal P}_{(0)}
\end{pmatrix}.
\lab{prs}
\ea
The number density ${\cal N}_{(0)}$, pressure $\mathcal{P}_{(0)}$,
and energy density ${\cal E}_{(0)}$
for the spinless and neutral classical massive particles 
read~\cite{Florkowski:2017ruc,Florkowski:1321594}
\beq
{\cal N}_{(0)} &=& \frac{T^3}{2\pi^2} z^2 \, K_2\left( z\right), \quad {\cal P}_{(0)} = T {\cal N}_{(0)}, \nn\\
{\cal E}_{(0)} &=& \frac{1}{2\pi^2} T^4 \, z ^2  \, 
 \left[z  K_{1} \left( z  \right) + 3 K_{2}\left( z \right) \right].
 \label{eneden}
\eeq
In the above equations $K_n(z)$ are the modified Bessel functions 
of the second kind~\cite{olver2010nist}
\begin{eqnarray}
 K_n(z) = \frac{z^n}{(2n + 1)!!} \int_1^\infty dx 
 (x^2 - 1)^{n -\frac{1}{2}} e^{-x z}.
\end{eqnarray}
The derivatives of $\mathcal{E}$ with respect to $\beta$ and
of $\mathcal{N}$ with respect to $\xi$ are
\begin{eqnarray}
 \left(\frac{\partial \mathcal{E}}{\partial \beta}\right)_\xi &=& 
 -\frac{2 m^3 T^2}{\pi^2} \cosh \xi \, 
 [z K_2(z) + 3K_3(z)],\nonumber\\
 \left(\frac{\partial \mathcal{N}}{\partial \xi}\right)_\beta &=& 
 \frac{2 m^2 T}{\pi^2} \cosh\xi\, K_2(z).
\end{eqnarray}
Taking into account that $(\partial \mathcal{E} / \partial T)_\xi = -\frac{1}{T^2} (\partial \mathcal{E} / \partial \beta)_\xi$, the functions $\mathcal{A}_1$ and $\mathcal{A}_3$ introduced 
in Eq.~\eqref{eq:A1&A3} can be readily computed:
\begin{eqnarray}
 \mathcal{A}_1 &=& \frac{4 \mathfrak{s}^2 m T^2}{3\pi^2} 
 \cosh\xi \left[K_3(z) + \frac{z}{2} K_2(z)\right], \nonumber\\
 \mathcal{A}_3 &=& -\frac{4 \mathfrak{s}^2 m T^2}{3\pi^2} 
 \cosh\xi \, K_3(z),
 \label{eq:A1&A3_MJ}
\end{eqnarray}
in agreement with the results reported in 
Ref.~\cite{Florkowski:2021wvk} for the ideal gas case. Substituting the above results in Eq.~\eqref{eq:cspin} gives Eq.~\eqref{eq:cspin_Kideal}.

We now discuss the asymptotic properties of the spin velocity $c_{\rm spin}$ in Eq.~\eqref{eq:cspin_Kideal} in the nonrelativistic and ultrarelativistic limits.
At large values of their argument, the modified Bessel functions admit the following asymptotic expansion \cite{olver2010nist}:
\begin{eqnarray}
 K_\nu(z) &=& \sqrt{\frac{\pi}{2z}} e^{-z} \sum_{k = 0}^\infty \frac{a_k(\nu)}{z^k}, \nonumber\\
 a_k(\nu) &=& \frac{\left(\frac{1}{2} - \nu\right)_k \left(\frac{1}{2} + \nu\right)_k}{(-2)^k k!}.
 \label{eq:K_largez_gen}
\end{eqnarray}
In particular, 
\begin{eqnarray}
 K_2(z) &=& \sqrt{\frac{\pi}{2z}} e^{-z} \left(1 + \frac{15}{8z} + \frac{105}{128z^2} + \dots \right), \nonumber\\
 K_3(z) &=& \sqrt{\frac{\pi}{2z}} e^{-z} \left(1 + \frac{35}{8z} + \frac{945}{128z^2} + \dots \right).
 \label{eq:K_largez}
\end{eqnarray}
From here, we obtain
\begin{eqnarray}
 c_{\rm spin}(z \gg 1) \simeq \frac{1}{\sqrt{2z}}.
 \label{eq:cspin_ideal_largez}
\end{eqnarray}
For small values of their argument, the modified Bessel functions of the 
second kind $K_n(z)$ of integer order $n$ admit the series representation
\cite{olver2010nist}
\begin{multline}
 K_n(z) = \frac{1}{2} \left(\frac{z}{2}\right)^{-n} 
 \sum_{k = 0}^{n-1} \frac{(n - k-1)!}{k!} \left(-\frac{z^2}{4}\right)^k \\
 + (-1)^{n+1} \ln \left(\frac{z}{2}\right) I_n(z) \\
 + \frac{(-1)^n }{2} \left(\frac{z}{2}\right)^n 
 \sum_{k = 0}^\infty [\psi(k+1) + \psi(n + k+ 1)] \frac{(z^2/4)^k}{k! (n+k)!},
 \label{eq:K_smallz_gen}
\end{multline}
where $\psi(z) = \Gamma'(z) / \Gamma(z)$ is the digamma function. 
The modified Bessel functions of the first kind $I_n(z)$ have the 
series representation 
\begin{eqnarray}
 I_n(z) = \left(\frac{z}{2}\right)^n \sum_{k = 0}^\infty \frac{(z^2 / 4)^k}{k!(n + k)!}.
\end{eqnarray}
Thus, the leading order contributions to $K_2(z)$ and $K_3(z)$ are 
given by the terms in the sum appearing on the first line of 
Eq.~\eqref{eq:K_smallz_gen},
\begin{align}
 K_2(z) =& \frac{2}{z^2} - \frac{1}{2} + O(z^2), &
 K_3(z) =& \frac{8}{z^3} - \frac{1}{z} + O(z).
 \label{eq:K_smallz}
\end{align}
Substituting the above into Eq.~\eqref{eq:cspin_Kideal} gives
\begin{align}
 c_{\rm spin}^2 (z \ll 1) \simeq& \frac{1}{4} 
 \left[1 - \frac{z^2}{8} + O(z^4)\right].
 \label{eq:cpsin_ideal_smallz}
\end{align}
\section{Fermi-Dirac gas}\label{app:FD}
%
The FD distribution is
\begin{eqnarray}
 f^{\sigma}_{\rm eq} = \frac{1}{e^{\beta p \cdot U - \sigma \xi} + 1}\,.
\end{eqnarray}
The charge density, energy density, and pressure can be computed as~\cite{Floerchinger:2015efa}
\begin{eqnarray}
 \begin{pmatrix}
  \mathcal{N} \\ 
  \mathcal{E} \\
  \mathcal{P}
 \end{pmatrix}&=& \frac{1}{\pi^2} \sum_\sigma
 \int_m^\infty dE\,p
 \begin{pmatrix}
  \sigma E \\
  E^2 \\
  \frac{1}{3} p^2
 \end{pmatrix}
 \frac{1}{e^{\beta E - \sigma \xi} + 1}.
 \label{eq:NEP_FD}
\end{eqnarray}
In the case when $\xi < z = \beta m$, the Fermi-Dirac factor $[e^{\beta E - \sigma \xi} + 1]^{-1}$ can be 
expanded as
\begin{eqnarray}
 \frac{1}{e^{\beta E - \sigma \xi} + 1} = \sum_{\ell = 1}^\infty (-1)^{\ell + 1} 
 e^{-\ell \beta E + \ell \sigma \xi}.\label{eq:FD_expansion}
\end{eqnarray}
This allows $\mathcal{N}$ and $\mathcal{E}$ to be computed as
\begin{align}
 \mathcal{N} =& \frac{2m^2 T}{\pi^2} \sum_{\ell = 1}^\infty 
 \frac{(-1)^{\ell + 1}}{\ell} \sinh(\ell \xi) K_2(\ell z),\\
 \mathcal{E} =& \frac{2m^2 T^2}{\pi^2} \sum_{\ell = 1}^\infty 
 \frac{(-1)^{\ell + 1}}{\ell^2} \cosh(\ell \xi) [\ell z K_1(\ell z) + 3 K_2(\ell z)],\nonumber
\end{align}
which has as its $\ell = 1$ contribution the result for the MJ statistics given in Eqs.~\eqref{prs} and \eqref{eneden}.

The derivatives of $\mathcal{E}$ and $\mathcal{N}$ with respect to $\beta$ and $\xi$, respectively, can be calculated as
\begin{eqnarray}
 \left(\frac{\partial \mathcal{E}}{\partial \beta}\right)_\xi &=& 
 -\frac{1}{\pi^2 \beta} \sum_{\sigma}
 \int_m^\infty dE\,E^2 
 \frac{3 p + \frac{E^2}{p}}{e^{\beta E - \sigma \xi} + 1},\nonumber\\
 \left(\frac{\partial \mathcal{N}}{\partial \xi}\right)_\beta &=& 
 \frac{1}{\pi^2 \beta} \sum_\sigma 
 \int_m^\infty dE
 \frac{p + \frac{E^2}{p}}{e^{\beta E - \sigma \xi} + 1}.
 \label{eq:d_EN_FD}
\end{eqnarray}
In the case when $|\xi| < z$, the above integrals can be computed using the method introduced in Eq.~\eqref{eq:FD_expansion}. This allows the functions $\mathcal{A}_1$ and $\mathcal{A}_3$ introduced in Eq.~\eqref{eq:A1&A3} to be expressed as
\begin{eqnarray}
 \mathcal{A}_1 &=& \frac{4 \mathfrak{s}^2 m T^2}{3\pi^2} 
 \sum_{\ell = 1}^\infty \frac{(-1)^{\ell + 1}}{\ell}
 \cosh(\ell \xi) \left[K_3(\ell z) + \frac{\ell z}{2} K_2(\ell z)\right], \nonumber\\
 \mathcal{A}_3 &=& -\frac{4 \mathfrak{s}^2 m T^2}{3\pi^2} 
 \sum_{\ell = 1}^\infty \frac{(-1)^{\ell + 1}}{\ell}
 \cosh(\ell \xi) K_3(\ell z),
 \label{eq:A1&A3_FD_smallxi}
\end{eqnarray}
where again the $\ell = 1$ term coincides with the expressions obtained in Eq.~\eqref{eq:A1&A3_MJ} for the MJ statistics. The above result is useful to derive the nonrelativistic and ultrarelativistic limits of $c_{\rm spin}$. In the latter case, the modified Bessel functions 
can be approximated by their large $z$ expansion, given in Eq.~\eqref{eq:K_largez}. In this case, the terms with $\ell > 1$ are penalized by the exponential function, $K_n(\ell z) \sim e^{-\ell z} / \sqrt{\ell z}$, such that the $\ell = 1$ term already provides a good approximation. For this reason, the value of $c_{\rm spin}$ corresponding to FD particles converges to the MJ one, given in 
Eq.~\eqref{eq:cspin_ideal_largez}. 

\begin{figure}
\centering
\includegraphics[width=\columnwidth]{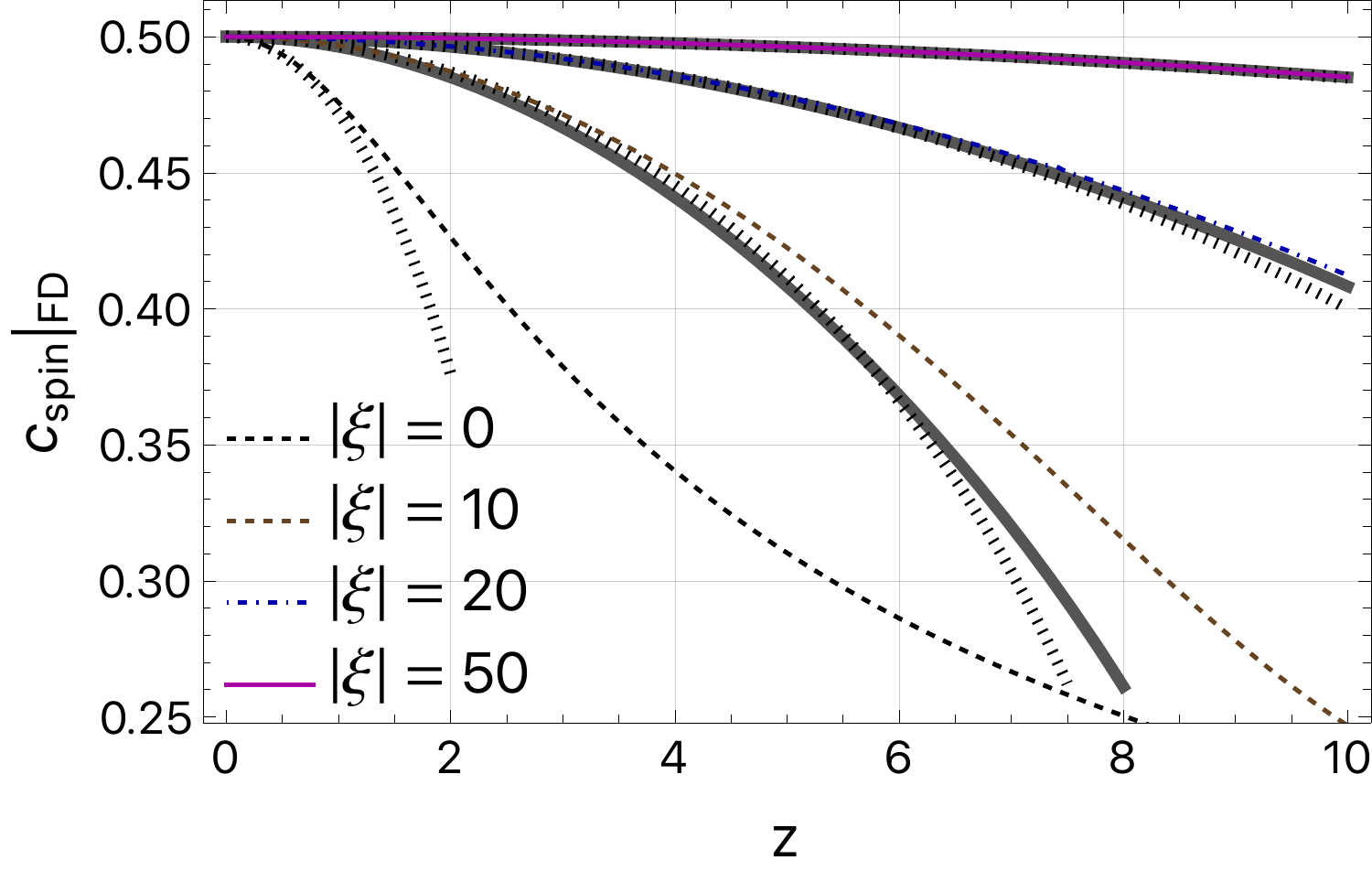}
\caption{Comparison between the numerical results for $c_{\rm spin}$ corresponding to FD statistics and the limits corresponding to the degenerate Fermi gas derived in Eq.~\eqref{eq:cspin_FD_deg} (thick gray lines, only when $|\xi| = 10$, $20$, and $50$) and to the small $z$ limit \eqref{eq:cspin_smallz} 
(dotted black lines).
\label{fig:degenerate_FD}
}
\end{figure}

In the relativistic limit, $z$ can be assumed to 
be small and the modified Bessel functions can be replaced by their asymptotic expansions in Eq.~\eqref{eq:K_smallz}. Denoting 
$S_n = \sum_{\ell = 1}^{\infty} \frac{(-1)^{\ell + 1}}{\ell^n} \cosh(\ell \xi)$,the functions $\mathcal{A}_1$ and $\mathcal{A}_3$ converge to
\begin{eqnarray}
 \mathcal{A}_1 &=& \frac{32 \mathfrak{s}^2 T^4}{3\pi^2 m^2} 
 [S_4 + O(z^4)],\nonumber\\
 \mathcal{A}_3 &=& -\frac{32 \mathfrak{s}^2 T^4}{3\pi^2 m^2} 
 \left[S_4 - \frac{z^2}{8} S_2 + O(z^4)\right].
\end{eqnarray}
Taking into account that $S_4 = \frac{1}{720}(7\pi^4 + 30\pi^2 \xi^2 + 15\xi^4)$and $S_2 = \frac{1}{12}(\pi^2 + 3\xi^2)$, we arrive at
\begin{eqnarray}
 c_{\rm spin}^2 &=& \frac{1}{4} \left[1 - \frac{z^2}{8} \frac{S_2}{S_4} + O(z^4)\right]\nonumber\\
 &=& \frac{1}{4} \left[1 - \frac{15z^2}{2\pi^2} \frac{1 + \frac{3\xi^2}{\pi^2}}
 {7 + 30 \frac{\xi^2}{\pi^2} + 15 \frac{\xi^4}{\pi^4}} + O(z^4)\right].
 \label{eq:cspin_FD_smallz}
\end{eqnarray}

Before ending this section, we discuss another interesting limit relevant 
for the FD statistics.
In the degenerate case ($T \rightarrow 0$ and $\mu > m$), we have
\begin{eqnarray}
 \mathcal{N} = \frac{p_F^3}{3\pi^2}, \,\,\,\,
 \mathcal{E} = \frac{1}{8\pi^2}
 \left[p_F \, \mu (p_F^2 + \mu^2) + 
 m^4 \ln \frac{m}{p_F + \mu}\right],\nn
\end{eqnarray}
where $p_F = \sqrt{\mu^2 - m^2}$. Since in the degenerate limit, $\mathcal{E} \equiv \mathcal{E}(\mu)$, we have 
$\left(\partial \mathcal{E} / \partial \beta\right)_\xi =-\frac{\mu}{\beta} (\partial \mathcal{E} / \partial \mu)$,
such that
\begin{eqnarray}
 \left(\frac{\partial \mathcal{E}}{\partial \beta}\right)_\xi = -\frac{\mu^3 \sqrt{\mu^2 - m^2}}{\pi^2 \beta}, \,\,\,\,
 \left(\frac{\partial \mathcal{N}}{\partial \xi}\right)_\beta = \frac{\mu \sqrt{\mu^2 - m^2}}{\pi^2 \beta}.~~~~~~~~~
\end{eqnarray}
This leads to the approximate formula
\begin{eqnarray}
 c_{\rm spin}^2 = \frac{1}{4} \frac{\xi^2 - z^2}{\xi^2 + z^2/2},
 \label{eq:cspin_FD_deg}
\end{eqnarray}
which is valid when $\xi \gg z$, see Fig.~\ref{fig:degenerate_FD} for the comparison between $c_{\rm spin}$ and the degenerate limit for the FD gas.
We can attempt to link $c_{\rm spin}$ to the sound speed for a degenerate 
gas. Taking into account the expression for the pressure,
\begin{eqnarray}
 \mathcal{P} = \frac{1}{24\pi^2}
 \left[p_F \mu (5p_F^2 - 3\mu^2) -
 3m^4 \ln \frac{m}{p_F + \mu}\right],
\end{eqnarray}
it can be shown that 
\begin{eqnarray}
 c_s^2 = \frac{d\mathcal{P}}{d\mathcal{E}} 
 = \frac{1}{3} - \frac{z^2}{3\xi^2}\,.
\end{eqnarray}
Comparing the above expression with Eq.~\eqref{eq:cspin_FD_deg}, the following relation can be established between the 
spin wave velocity and the sound velocity in the degenerate limit for the FD gas
\beq
c_{\rm spin}^2 = \frac{c_s^2/2}{1 - c_s^2}.
\eeq
\bibliography{pv_ref}{}
\bibliographystyle{utphys}
\end{document}

%% file: commands.tex
\newcommand{\e}{{\rm e}}

\newcommand{\n}{\nonumber}

\def\be{\begin{equation}}
	\def\ee{\end{equation}}
	\newcommand{\eel}{\end{eqnarray}}
\def\barr{\begin{array}}
	\def\earr{\end{array}}
\def\beq{\begin{eqnarray}}
	\def\eeq{\end{eqnarray}}
\def\bfig{\begin{figure}}
	\def\efig{\end{figure}}
\newcommand{\bea}{\begin{eqnarray}}
	\newcommand{\eea}{\end{eqnarray}}

\def\LB{\left(}
\def\RB{\right)}

\newcommand{\nn}{\nonumber}

\newcommand{\p}{\partial}
 
\def\a{\alpha}
\def\b{\beta}
\def\g{\gamma}
\def\d{\delta}

\def\LR{\left(} 
\def\RR{\right)}

\def\HP{\hphantom{\alpha}} 

\def\half{\frac{1}{2}}

\newcommand{\lab}[1]{\label{#1}}
\def\nn{\nonumber}

\def\av{{\boldsymbol a}}
\def\bv{{\boldsymbol b}}

\def\omnU{\omega^{\mu\nu}}

\def\be{\begin{equation}}
\def\ee{\end{equation}}
\def\ba{\begin{eqnarray}}
\def\ea{\end{eqnarray}}   

\def\a{\alpha}
\def\b{\beta}
\def\g{\gamma}
\def\d{\delta}

\def\LR{\left(} 
\def\RR{\right)}

\def\half{\frac{1}{2}}

\def\av{{\boldsymbol a}}
\def\bv{{\boldsymbol b}}

\def\omnU{\omega^{\mu\nu}}

\def\half{\frac{1}{2}}

\def\n0{n_{(0)}}
\def\e0{\varepsilon_{(0)}}
\def\P0{P_{(0)}}

\def\bv{{\boldsymbol b}}
